# Towards the Theory of Isotope Effect of the London Penetration Depth in Cuprates

I.E. Lyubin[1], M.V. Eremin[1], I.M. Eremin[1,2], H. Keller[3]

[1] Kazan State University, 420008 Kazan, Russian Federation
[2] Max-Plank Institute für Physik komplexer Systems, Dresden, D-01187, Germany
[3] Physik-Institute, Universität Zurich, CH-8057, Switzerland

In this short note, we discuss the expressions for the superfluid density in both weak- and strong-coupling approaches. We further perform the numerical calculations of temperature and isotope composition dependencies for a number of High-$T_c$ compounds. We employ the tight-binding parameters and the corresponding Fermi surfaces in accordance to the available photoemission data.

In a weak-coupling approach the kinetic energy of carriers in the external vector potential $A$ is written as $\left( p - \frac{e}{c} A \right)^2 / 2m$. It this case the superfluid density is calculated as follow [1]

$$\lambda_L^{-2}(T) = \lambda_L^{-2}(0) \left[ 1 + 2 \int_\Delta^\infty \left( \frac{\partial f(E)}{\partial E} \right) \frac{E}{(E^2 - |\Delta_k|^2)^{1/2}} dE \right]. \qquad (1)$$

Here $E_k = \sqrt{(\varepsilon_k - \mu)^2 + |\Delta_k|^2}$ are the energy of Bogolyubov's quasiparticles and $\varepsilon_k = t_k \approx (\hbar k)^2 / 2m$. It is well known, that according to ARPES data (see ref. [2]), the energy dispersion of the charge carries in layered cuprates is well fitted using the expression

$$\varepsilon_k = 2t_1(\cos k_x a + \cos k_y a) + 2t_2 \cos k_x a \cos k_y a + 2t_3 (\cos 2k_x a + \cos 2k_y a) + \ldots \qquad (2)$$

In practice this means that there exist substantial deviations of the energy dispersion from the simple parabolic form which Eq. (1) assumes. In such case, the presence of an external vector potential $A$ yields the following renormalization of the hopping integrals

$$t_{ij} \to t_{lj} \exp\left[ -i \frac{e}{\hbar c} \int_{R_l}^{R_j} \mathbf{A} d\mathbf{s} \right] \approx t_{lj} \exp\left\{ -i \frac{e}{\hbar c} A_x R_{jl}^x \right\}. \qquad (3)$$

Calculating the kinetic energy variation in a magnetic field and comparing the result with classical formula $\delta H_{kin}^{(1)} = -\frac{1}{c} A_q^x j_x(-q) + h.c.$ we deduce the expression for the electrical current density

$$j(q) = -\frac{e}{2\hbar} \sum_k \left[ \frac{\partial \varepsilon_k}{\partial k_x} + \frac{\partial \varepsilon_k}{\partial (k_x + q_x)} \right] a_{k,\sigma}^+ a_{k+q,\sigma}. \qquad (4)$$

Then using the London's equation $j(q=0) = -\frac{c}{4\pi\lambda^2} A_{q=0}$ we arrive to expression for the superfluid density

$$\frac{1}{\lambda^2} = \frac{e^2}{c\hbar^2} \sum_k \frac{\partial \varepsilon_k}{\partial k_x} \left[ \frac{|\Delta_k|^2}{E_k^2} \frac{\partial \varepsilon_k}{\partial k_x} - \frac{(\varepsilon_k - \mu)}{2E_k^2} \frac{\partial |\Delta_k|^2}{\partial k_x} \right] \left[ \frac{1}{E_k} - \frac{\partial}{\partial E_k} \right] \tanh\left( \frac{E_k}{2k_B T} \right). \qquad (5)$$

This agrees well with the results obtained previously in Ref. [3, 4] with one important difference. In particular, Eq. (5) contains the absolute square of the gap function, $|\Delta_k|^2$ and not $\Delta_k^2$ as it was suggested [3, 4]. This consequently results in a number of differences between Eq. (5) and Eq. (1) that we would like to point out. In contrast to Eq. (1) the paramagnetic current contribution to Eq. (4) strongly depends on temperature. In (1) this is the first term in brackets and, as one can easily verify is temperature independent. In addition, there are also differences in the behavior of the diamagnetic contributions.

Part of our numerical calculations is displayed in Fig. 1. The superconducting gap has been introduced phenomenologically as in Ref. [5]

$$\Delta_k(T) = \frac{\Delta_0}{2}\left(\cos k_x a - \cos k_y a\right)\tanh\left(1.76\sqrt{T_c/T - 1}\right) \qquad (6)$$

The isotope coefficient $\alpha_\Delta = 0.0248$ [$^{18}\Delta_0 = {}^{16}\Delta_0\left(1 - \alpha_\Delta \frac{\Delta m}{m}\right)$] is taken according to experimental data [6] and employing Eq. (6). Calculations were done assuming that all hopping integrals [7] change under substitution of $^{18}O$ by $^{16}O$ following the polaronic renormalization $^{18}t = {}^{16}t\left(1 - \alpha_t \frac{\Delta m}{m}\right)$. The isotope coefficient is extracted to be $\alpha_t = 0.3545$.

We further note that the isotope coefficients on the superconducting gap, $\alpha_\Delta$ and on the hopping integral, $\alpha_t$, have to be consistent. In other words, assuming the isotope effect on $\alpha_t$ and substituting it in the gap equation for the superconducting gap we have to find $\alpha_\Delta$ that is close to the value extracted from experiment. In order to see this we solve the BSC integral equation

$$\Delta(k) = \frac{1}{N}\sum V(k-k')\frac{\Delta(k')}{E_{k'}}\tanh\left(\frac{E_{k'}}{2k_B T}\right), \qquad (7)$$

where the pairing potential acts between nearest neighbors $V(q) = 2J\left(\cos q_x a + \cos q_y a\right)$. Assuming $V(q)$ is independent on the oxygen mass, the calculated isotope coefficient $\alpha_\Delta$ was found small and *negative* as shown in Fig. 2. This is in contrast to the experimentally extracted value. This suggests that electron-phonon interaction entering $V(q)$ is non-negligible and has to be taken into account. In particular, we expect that including electron-phonon contribution in addition to the density-density [8] and superexchange [9] parts to the pairing potential will reverse $\alpha_\Delta$ to a *positive* value. Therefore we conclude that (i) the main reason for the isotope effect originates from renormalisation of hopping parameters; (ii) isotope effect on penetration depth, related to the superconducting energy gap shift is relatively small, and (iii) the role of electron-phonon mediated interaction entering the pairing interaction is more important as usually appreciated.

This work is partially supported by the Swiss National Science Foundation, Grant IB7420-110784 and the Russian Foundation for Basic Research, Grant N 06-0217197-a.

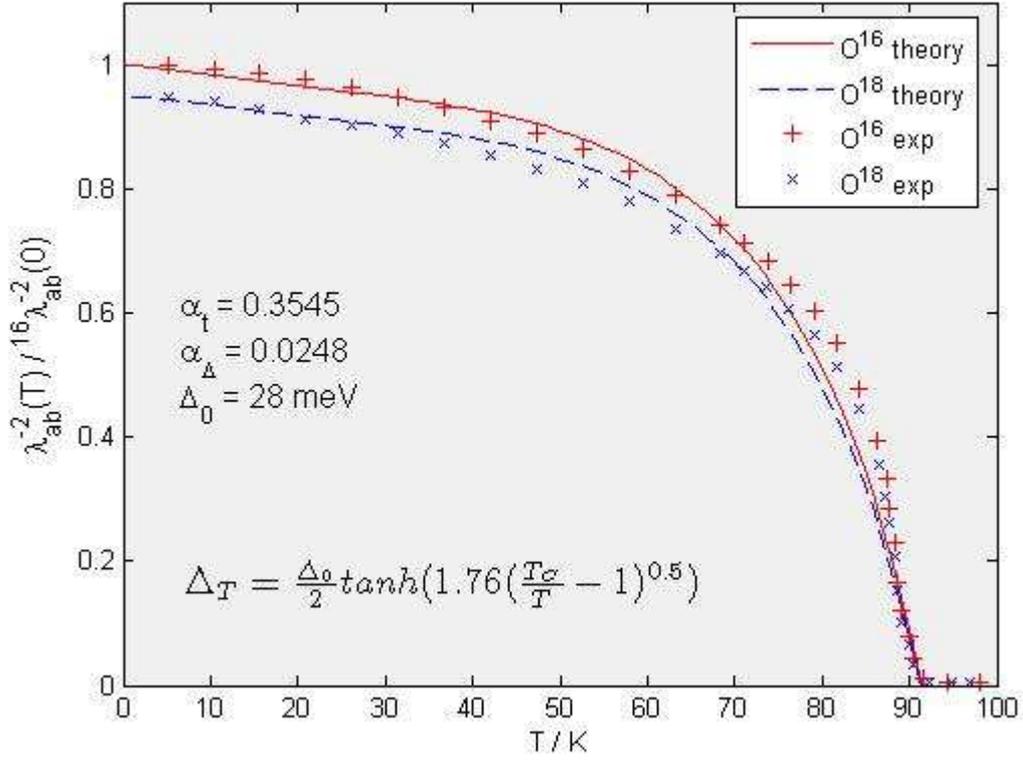

**Fig. 1.** Calculated temperature evolution of the penetration depth in $YBa_2Cu_3O_{6.98}$. Symbols are experimental points from Ref. [6], solid curves denote our calculations as obtained from Eq. (5). The hopping parameters were taken as in Ref. [7].

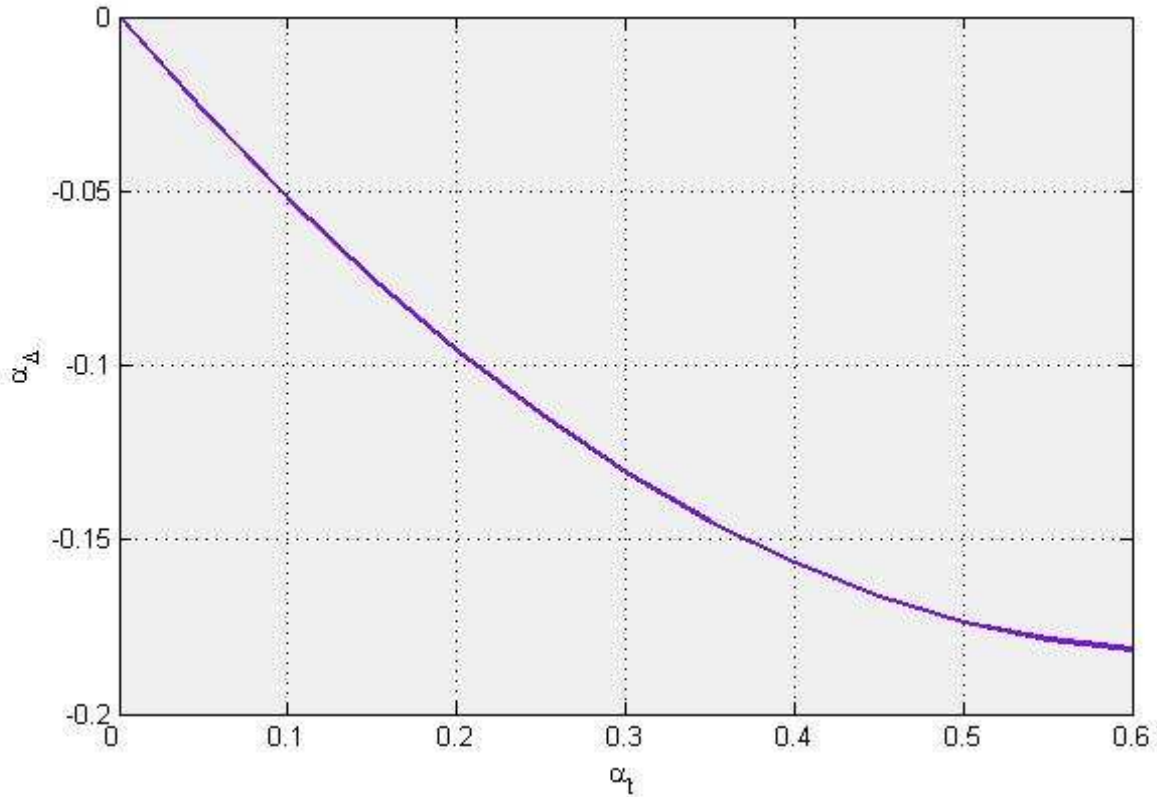

**Fig. 2.** Isotope coefficient for the superconducting gap versus $\alpha_t$ assuming paring potential in Eq. (7) is independent on the isotope mass.